\documentclass[a4paper]{jpconf}

\usepackage{color}

\usepackage[utf8]{inputenc}
\usepackage{graphicx}
\newcommand{\beq}{\begin{equation}}
\newcommand{\beqa}{\begin{eqnarray}}
\newcommand{\eeq}{\end{equation}}
\newcommand{\eeqa}{\end{eqnarray}}

\def\vec#1{\ensuremath{\mathchoice{\mbox{\boldmath$\displaystyle#1$}}
{\mbox{\boldmath$\textstyle#1$}}
{\mbox{\boldmath$\scriptstyle#1$}}
{\mbox{\boldmath$\scriptscriptstyle#1$}}}}

\begin{document}
\title{MHD simulations of resistive viscous accretion disk around
millisecond pulsar }

\author{M \v{C}emelji\'{c}, V Parthasarathy and W Klu\'{z}niak}

\address{Nicolaus Copernicus Astronomical Center, ul. Bartycka 18, PL-00-716
Warsaw, Poland}

\ead{miki@camk.edu.pl}

\begin{abstract}
We perform MHD simulations of a thin resistive and viscous accretion disk
around a neutron star with the surface dipolar magnetic field of 10$^8$~Gauss.
The system evolution is followed during the interval of 500 millisecond pulsar
rotations. Matter is accreted through a stable accretion column from the disk 
onto the star. We also show propagation of the stellar wind through the corona.
Analysis of the mass accretion flux and torques on the star shows that the
disk reaches the quasi-stationary state.
\end{abstract}

\begin{figure}[h]
\begin{center}  
\includegraphics[width=24pc]{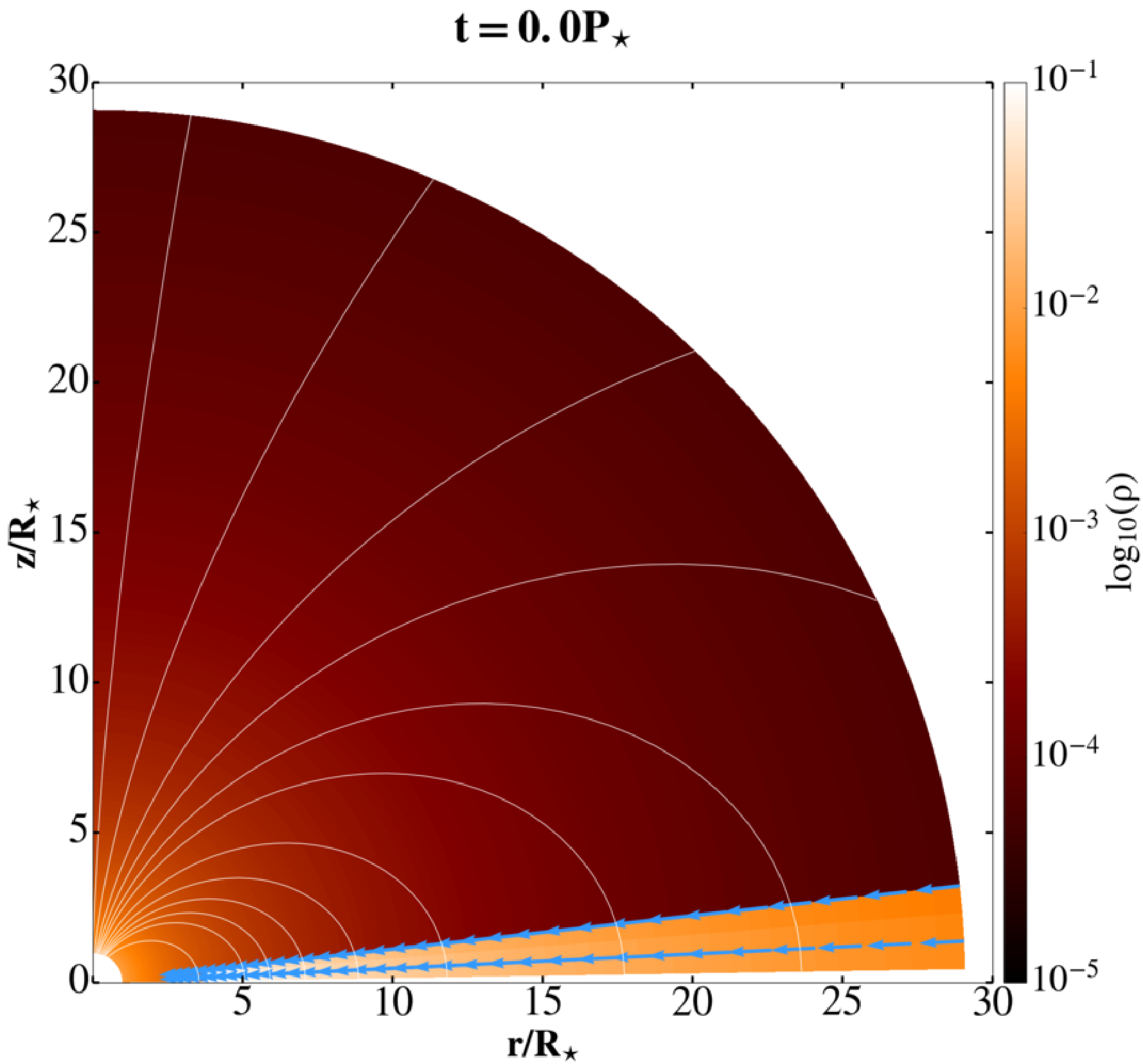}
\includegraphics[width=24pc]{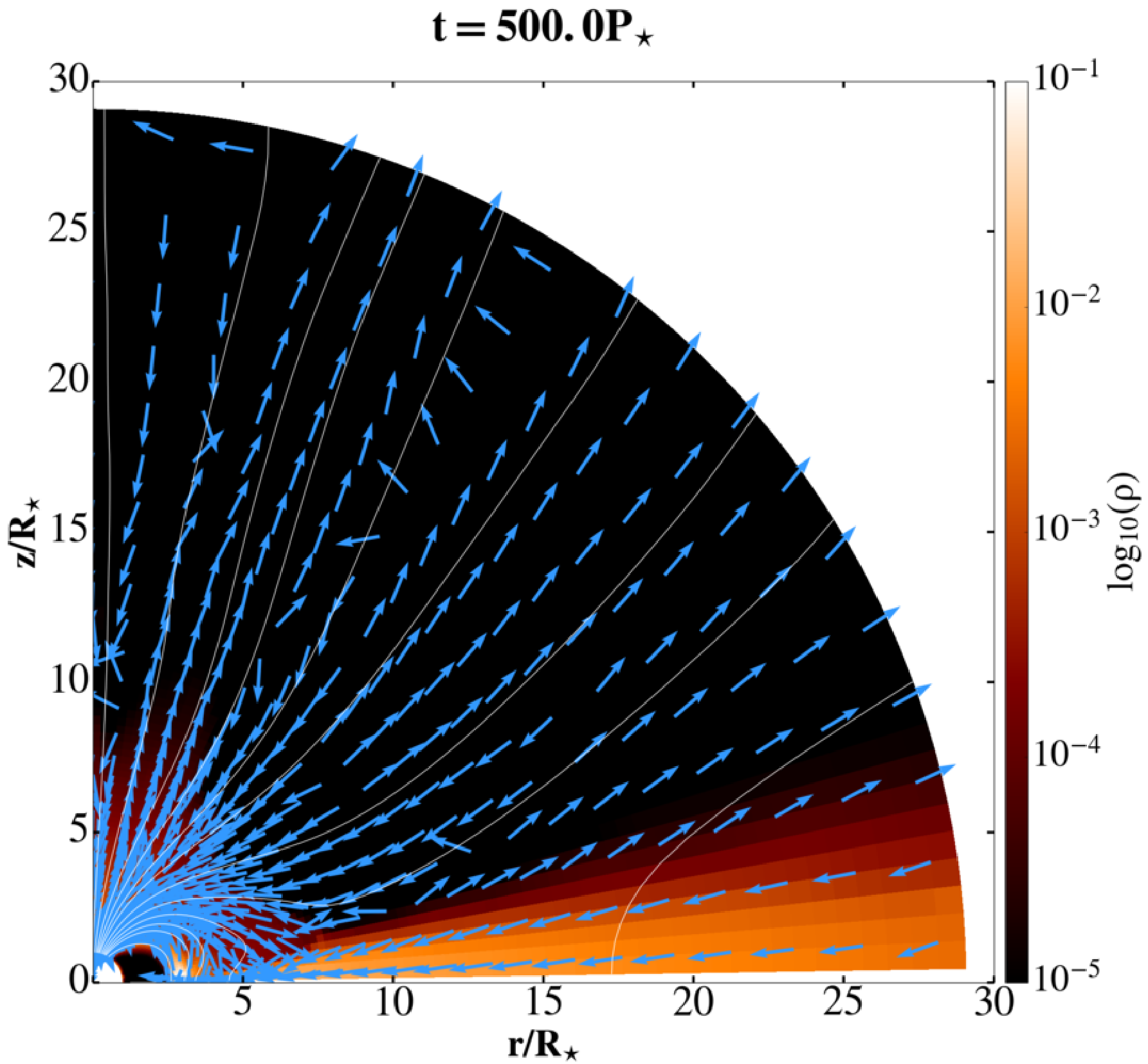}   
\caption{Simulated matter density distribution is shown with the logarithmic
color grading. The initial setup is in the top panel and the system
after 500 pulsar rotations is shown in the bottom panel. Magnetic field
lines are plotted by solid lines and the poloidal velocity distribution is
displayed by blue vectors (that are not normalized).}
\end{center}
\label{fig1}
\end{figure}

\begin{figure}[h]
\begin{center}  
\includegraphics[width=30pc]{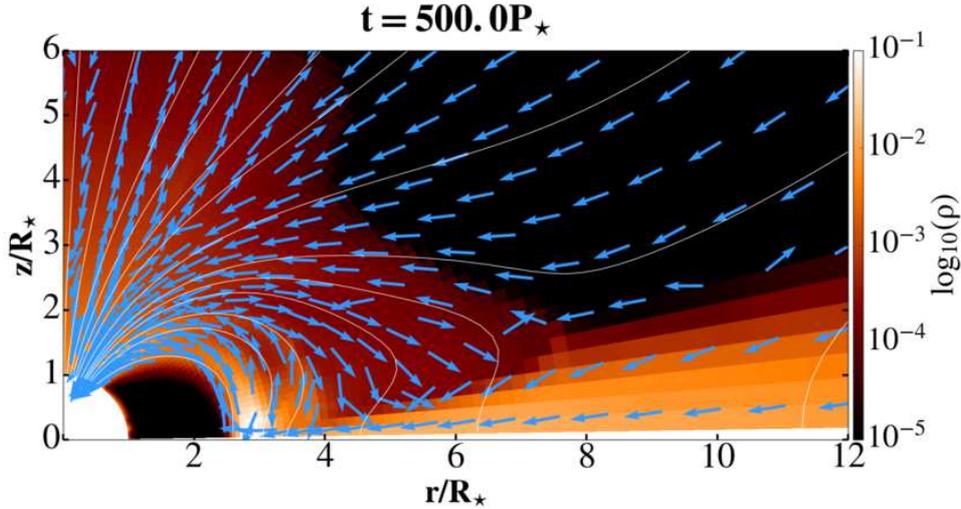}
\caption{Central part of the system after 500 pulsar rotations is presented
to visualize the accretion column and the magnetic field lines connected to
the disk beyond the corotation radius $R_{\mathrm cor}=4.65~R_*$.}
\end{center}
\label{fig2}
\end{figure}

\begin{figure}[h]
\begin{center}  
\includegraphics[width=18.5pc]{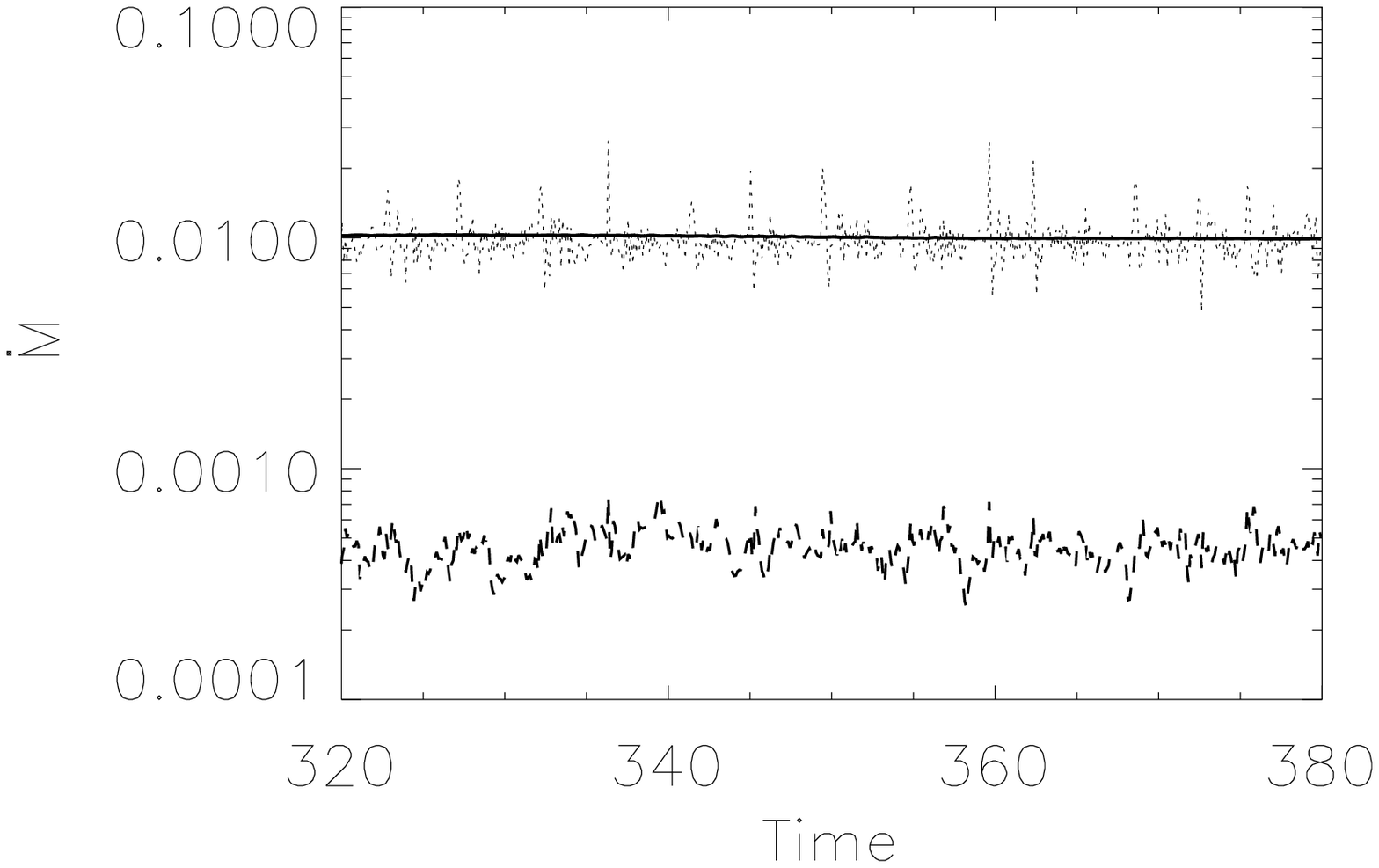}
\includegraphics[width=18.5pc]{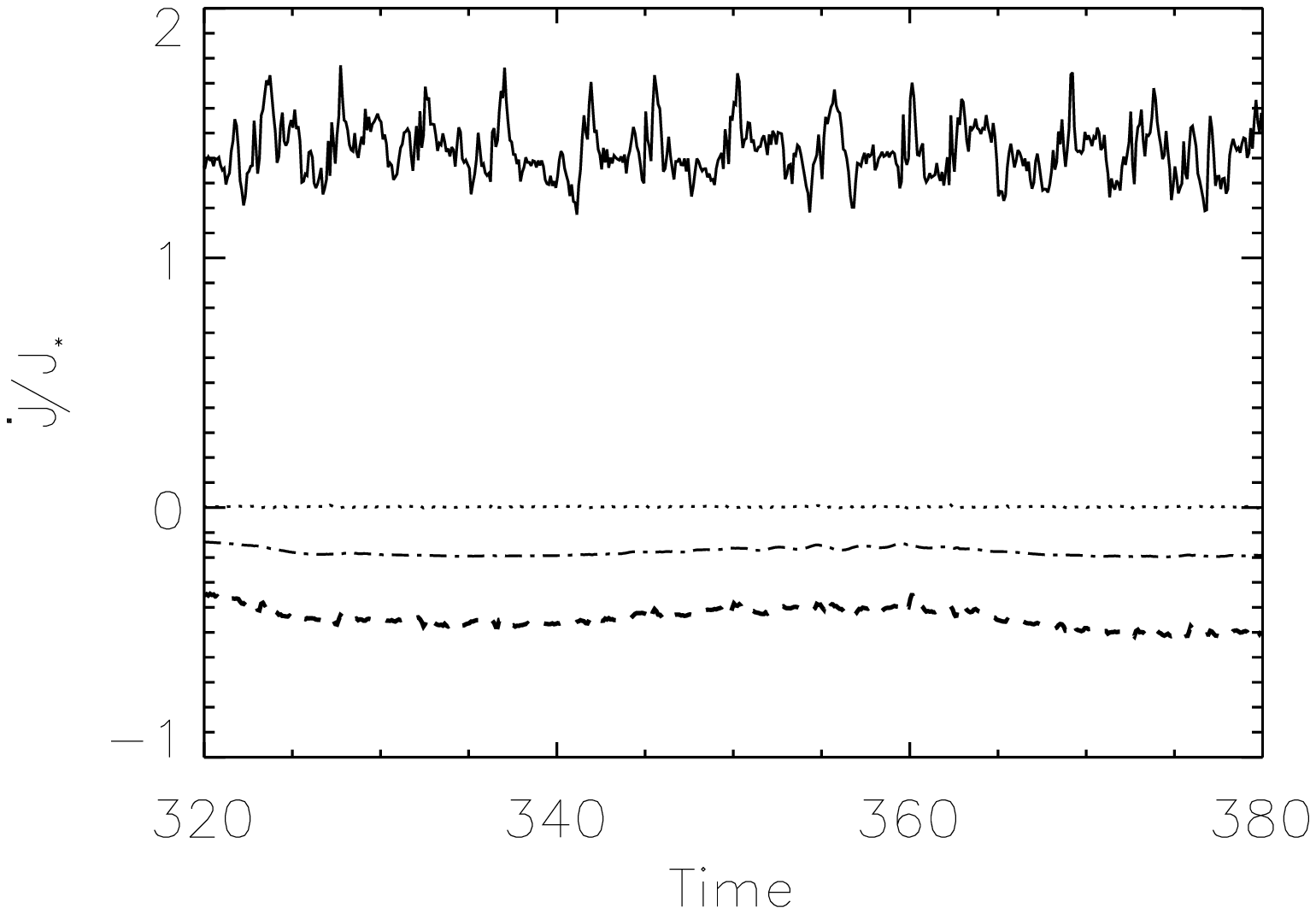}   
\caption{The left panel shows the time dependence of the mass flux.
The mass flux is measured in the code units
$M_0=\rho_0 R_*^2 v_{\mathrm K,*}$, and the time is measured in
the pulsar rotation periods. The time interval during the
quasi-stationary state is shown. Solid line shows the mass flux
through the disk at $R=12R_*$. This mass flux is distributed
onto the star (dotted line) and into the stellar wind (thin dashed line).
Right panel shows the torque acting on the stellar surface in the units
of the stellar angular momentum for the same time interval. 
Dotted line shows the kinetic torque which is negligible in this case
and dot-dashed line shows the magnetic torque on the star produced by
the stellar wind. Solid and dashed lines show the magnetic torque
acting on the star surface from the field lines below and beyond
the corotation radius. Positive torque spins-up and negative
spins-down the star.}
\end{center}
\label{fig3}
\end{figure}

\section{Introduction}
In the interaction of a neutron star (NS) with its close companion star,
an accretion disk is formed  around the NS. Properties of a binary system
depend on the type of companion star, the neutron star mass and the magnetic
field strength and geometry. Klu\'{z}niak \& Kita suggested a
hydro-dynamical model for the accretion disk [1], with viscosity
parameterized by Shakura \& Sunyaev $\alpha$-prescription [2].
We extend this model to the non-ideal MHD, and include the magnetosphere
in the innermost part of a star-disk system. One example of such object is
a millisecond pulsar: $M=1.4M_\odot$, $R\sim 10$~km, $B\sim 10^8$~Gauss,
$P=0.05$~sec (5~msec), $\rho_0=4.62\times 10^{-6}$~g/cm$^3$,
$\dot{M}_0=10^{-9} M_\odot$/yr.

\section{Numerical setup}
We use the PLUTO v.4.1 code [3,4] in spherical coordinates with logarithmically stretched grid in
radial direction to perform axisymmetric 2D star-disk simulations in
$\Theta=[0,\pi/2]$ half-plane. The resolution is $R\times \Theta = [109\times
50]$ grid cells, with the maximal radius of 30 stellar radii. Following [5]
we set up the disk as in [1] assuming that the corona is in a hydrostatic
equilibrium. The viscosity and the resistivity are
parameterized  as $\alpha c^2/\Omega$ (Shakura \& Sunyaev), where $c$ is
the sound speed, $\Omega$ is the Keplerian speed and $\alpha$ is a free
parameter, which values are between 0 and 1. We use a split-field method in
which only changes from the initial stellar magnetic field evolve in time
while the initial stellar field is held constant. Our initial setup is shown
in figure~1.

The equations we solve using the PLUTO code are: 
\beqa
\nonumber \frac{\partial \rho}{\partial t} + \nabla \cdot (\rho \vec{v})=0 ,\
\frac{\partial\rho\vec{v}}{\partial
t}+\nabla\cdot\left[\rho\vec{v}\vec{v}+\left(P+\frac{\vec{B}\cdot\vec{B}}{8\pi}\vec{I}\right)-\frac{\vec{B}\vec{B}}{4\pi}-\vec{\tau}\right]=-\rho\nabla\Psi_G\\
\nonumber
\frac{\partial E}{\partial
t}+\nabla\cdot\left[\left(E+P+\frac{\vec{B}\cdot\vec{B}}{8\pi}\right)\cdot\vec{v}-\frac{(\vec{v}\cdot\vec{B})\vec{B}}{4\pi}\right]=-\rho\nabla\Psi_G\cdot\vec{v}\\
\nonumber
{\mathrm with}\quad
E=\frac{P}{\gamma-1}+\rho\frac{\vec{v}^2}{2}+\frac{\vec{B}^2}{8\pi}\quad {\mathrm and}
\quad\frac{\partial\vec{B} }{\partial t}+\nabla\times(\vec{B}\times\vec{v}-\eta_m\vec{J})=0 \
,
\eeqa
where $\eta_m$ and $\tau$ are the resistivity and the viscous stress tensor
respectively. Following [5] we remove the Ohmic and the viscous heating
terms in the PLUTO energy equation to prevent the thermal thickening of the
accretion disk. The resistive and the viscous terms are still present in the
momentum and the induction equations. In the boundary conditions we assume the
stellar surface to be a rotating perfect conductor so the electric field is
zero in the stellar reference frame and the flow velocity is parallel to the
magnetic field. We additionally prescribe the rotation of the matter atop the
star to match the magnetic field evolution.

\section{Results}
We obtain long-lasting solutions for the millisecond pulsar as shown
in figure~1. In figure~2 we show a zoom into this solution to show the
accretion column and magnetospheric region in more detail.

Mass fluxes and torques obtained in our star-disk interaction simulation
are shown in figure~3. Most of the mass from the disk is accreted onto the
star and only about 1/100 of it goes into the stellar wind. The torque of
mass infalling onto the star depends on the origin of the mass: when it
comes from the region beyond the corotation radius it slows down the star
and when it comes from the region below the corotation radius it spins-up
the star.

\section{Conclusions}
We present the preliminary results of the long-lasting numerical simulations
of a star-disk system with magnetospheric interaction in the case of
a millisecond pulsar. The disk in our viscous and resistive MHD simulations
reaches the quasi-stationary state. In further work we will investigate
properties and stability of such a disk. To completely address the stability
of the disk and the accretion column the 3D simulations should also be
performed.

\ack
We thank A. Mignone and his team of contributors for the possibility to use
the PLUTO code, especially C. Zanni for help with modifications. M\v{C}
thanks to ASIAA/TIARA, Taiwan for the possibility to use their HPC Linux clusters.

\smallskip

\end{document}